# Linearity Limits of Biased 1337 Trap Detectors

*P. Balling, P. Křen: Czech Metrological Institute (CMI),*

*V botanice 4, 150 72 Prague 5, Czech Republic*

## Abstract

The upper power limit of linear response of light trap detectors was recently measured [2,3]. We have completed this measurement with test of traps with bias voltage at several visible wavelengths using silicon photodiodes Hamamatsu S1337-1010 and made a brief test of S5227-1010. Bias extends the linearity limit by factor of more than 10 for very narrow beams and more than 30 for wide beams [5]. No irreversible changes were detected even for the highest irradiance of 33 W/cm$^2$ at 406nm. Here we present measurement of minimal bias voltage necessary for 99%, 99.8% and 99.95% linearity for several beam sizes.

## 1. Introduction

Silicon trap detectors [1] are widely used as transfer standards in radiometry. They have
- simple <u>spectral sensitivity</u> (external quantum efficiency >0.99 electron per incident photon for wavelengths 440nm-950nm)
- excellent <u>homogeneity</u> of response across the aperture (variations less than 0,1%)
- <u>low dark current</u> (<1nA)
- and nice <u>linear response</u> – the photocurrent is proportional to the incident power for photocurrents 1 nA to 0.4 mA for unbiased traps and 1 nA to 25 mA for bias of 10V. The upper power linearity limit is not a function of wavelength, but function of photocurrent and beam diameter [2,5].

The above ranges for linearity response are valid for Gaussian beam diameters larger than 2mm. For more narrow beams the upper power limit of linear response decreases. Surprisingly, the nonlinearity is not a simple linear function of irradiance (inverse-proportional to square of diameter).

Here we present measurement of bias voltage necessary for 99%, 99.8% and 99.95% linearity for Gaussian beam diameters (~2mm, ~0.6mm, ~0.2mm, measured by a beam analyzer).

## 2. Influence of bias voltage
### 2.1 Bias Influence on dark current
Some dark current is always present in the signal of light trap even without bias voltage (typical 300pA±100pA for three diode trap [4]). The dark current increases according to shunt resistance of 20 GΩ (represents the effective shunt resistance of three Hamamatsu S1337-1010 diodes connected in parallel)[5]. This small increase of the dark current does not influence the precision of measurement - we have not observed any increase of noise and the stable value is subtracted from measured signal as a part of background.

In our tests all Si photodiodes (1337 and 5227) have survived bias voltage of 30V, but only 5V is guaranteed by manufacturer. We do not recommend exceeding guaranteed bias in the case of GaAsP photodiodes (Hamamatsu G2119) – it can cause some irreversible change and increase the dark current.

### 2.2 Bias Influence on measured signal
Figure 1 shows the dependences of measured signal on the bias voltage. The behaviour could be divided into three cases (areas):
a) the irradiance is low enough (within the unbiased-linearity limits[1])
b) the irradiance is above that limit - some non-linearity starts to be detectable for no or insufficient bias.
c) sufficient bias voltage for actual irradiance is applied

In the areas a) and c) the photocurrent remains constant for increasing bias (providing that nearly negligible dark current is subtracted), the absolute

---
[1] corresponds to photocurrents of about 0.4 mA for wide beam

sensitivity of photodiode or trap detector remains constant.

In the area b) the detector shall not be used. The photocurrent raises up with bias (up to the value corresponding to linear response in area c)) and even with time: in the period of about minute photocurrent increases by up to 10% but never reaches the correct value of linear response (see arrows in Fig.1).

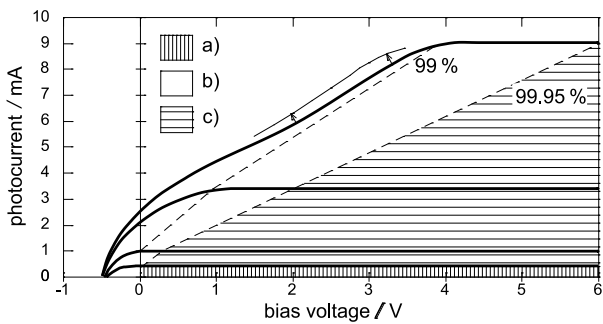

**Figure 1.** Photocurrent versus bias voltage for single diode (633 nm, 2 mm spot) for four different irradiance levels. In the areas below the dashed lines the linearity is better than 99 % and 99.95 %, respectively.

*2.3 Minimal necessary bias*

For linear response of traps it is important to know what bias we need to apply. Minimal necessary bias is function of both total power and beam diameter, but not simple linear function of irradiance as it was shown already in [3] for low nonlinearity.

We have found, that the minimal necessary bias voltage $U_{Bmin}$ for 99% linearity is very similar (Fig 2a and Table 1) for all beam diameters investigated and is proportional to $I_p$ (the photocurrent to be detected):

$$U_{Bmin}(99\%) \approx 300\Omega \cdot I_p - 0.5V \quad (1)$$

Precise estimation of bias voltage necessary for 100% linearity is more difficult. This minimal voltage strongly depends on the beam diameter – the smaller beam cross section the higher the necessary voltage (Tab 1 and Fig. 2).

$U_{Bmin}$ is fairly linear function of $I_p$ in the case of photocurrents above 2mA, so for certain linearity and beam diameter it can be described as follows:

$$U_{Bmin} = R \cdot I_p - U_{Bmin0} \quad (2)$$

The approximate values we have found are in Table 1.

The values R from Table 1 are presented also in Figure 3. The minimal necessary voltage is function of

- total power - proportional (for $I_p$ above 2mA)
- beam diameter $\underline{d}$ - inverse proportional (not inverse proportional to beam cross section but to beam diameter !)

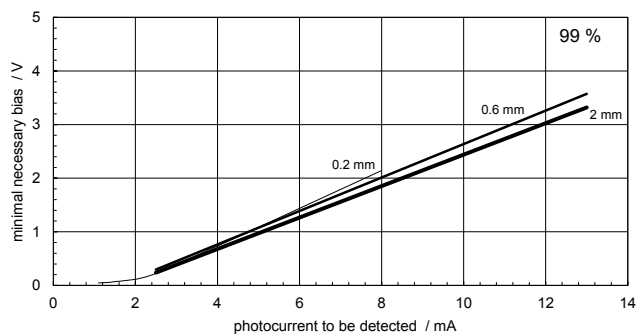

**Figure 2a.** Minimum necessary bias as a function of photocurrent $I_p$ for different beam diameters and 99 % linearity.

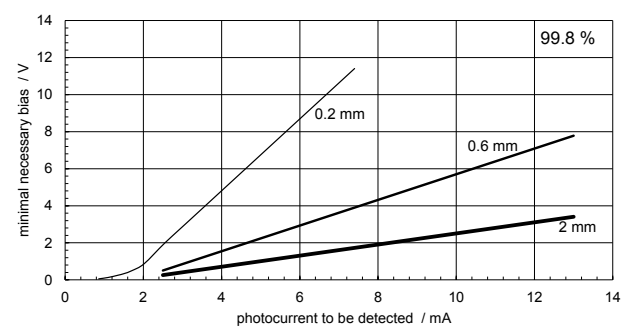

**Figure 2b.** Minimum necessary bias as a function of photocurrent $I_p$ for different beam diameters and 99.8 % linearity.

It is well known that photodiodes S1227-1010 have better linearity compared to S1337-1010. We have learned, that also the linearity of photodiodes S5227-1010 is very good (necessary bias is lower). The values of coefficients are in Table 2.

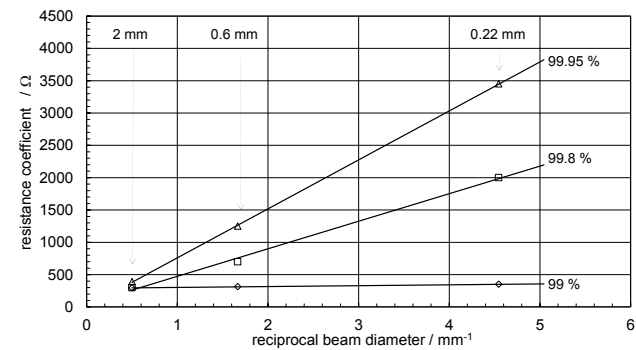

**Figure 3.** Coefficients *R* for different beam diameters and linearities. See (2) and Table 1.

**Table 1.** Coefficients *R* and $U_{Bmin0}$ for $I_p > 2$mA, for various Gaussian beam diameter and linearity (for three diode trap **1337**)

| linearity | 99% | | | 99.8% | | | 99.95% | | |
|---|---|---|---|---|---|---|---|---|---|
| diam. [mm] | 2 | 0.6 | 0.2 | 2 | 0.6 | 0.2 | 2 | 0.6 | 0.2 |
| R [Ω] | 296 | 312 | 380 | 300 | 700 | 2200 | 390 | 1250 | 3800 |
| $U_{Bmin0}$ [V] | 0.44 | 0.48 | 0.65 | 0.49 | 1.2 | 3.0 | 0.44 | 1.7 | 2.6 |

**Table 2.** Coefficients *R* and $U_{Bmin0}$ for $I_p > 5$mA, for various Gaussian beam diameter and linearity (for three diode trap **5227**)

| linearity | 99% | | | 99.8% | | |
|---|---|---|---|---|---|---|
| diam. [mm] | 1 | 0.35 | 0.2 | 1 | 0.35 | 0.2 |
| R [Ω] | 81 | 86 | 109 | 85 | 86 | 111 |
| $U_{Bmin0}$ [V] | 0.61 | 0.50 | 0.59 | 0.60 | 0.43 | 0.53 |

### 3. Practical application

As a supply of bias voltage it is convenient to use a battery, because of small size and elimination of possible problems with earthing or noise (we use rechargeable 4 x alkaline cell 1.5 V mechanically connected to trap). It enables us to precisely measure the power of eg. 2mm beams up to photocurrent of 14mA (28mW at 633nm). Our opinion is that this way is easier and more precise than the use of calibrated 5% filter and unbiased trap.

For to check the precision of measurement of unknown (narrow) beam, one can make a quick test – to decrease the bias voltage (e.g. by 50%) and set it back. If no change in photocurrent appears the value is correct. If some change in photocurrent appears, the error (difference of $I_p$ and photocurrent measured with higher bias) is definitely less than the difference between values measured in test[2]. If higher precision is needed, it is necessary to increase the bias and repeat the test.

The current to voltage converter is not needed while measuring the photocurrent of biased trap but the ampere-meter could be used directly (the voltage decrease on ampere-meter input (about 0.1V) is negligible compare to bias).

### 4. Uncertainty

The uncertainty of Gaussian beam diameter measurement was up to 20%.

The uncertainty of estimation of the minimal necessary bias in one particular set-up is quite good in the case of 99% linearity (better than 0.05V) but it is much worse in the case of 99.95% linearity, where we try to measure the change close to laser power instability (here the uncertainty raises up to several volts for intense and pointed beams). This uncertainty in one set-up then should be expanded due to uncertainty of beam diameter measurement, which is negligible for large beams but relatively large for pointed ones.

We do apologise that we don't state the uncertainty of every value we have measured or derived. We think that in this case it would make the tables and graphs even more confused. The purpose of this article is rather to illustrate the principle and to mark the areas where it is safe to work providing one does not approach too close to the limits.

### 5. Conclusions

Biased Si trap detectors can be used to extend the power range of applications as compared to the unbiased ones.

Approximate formulas were found for estimation of bias necessary for precise measurement of intense and pointed beams.

For beam diameter of 2 mm, the trap response remains linear within 0.05% (the uncertainty of measurement) for photocurrents up to 14 mA (corresponding to the optical power of 23mW at 752 nm or 43mW at 406.7 nm) for traps equipped with 1337 photodiodes and bias of 6V. For 5227 photodiodes the limits are likely to be about at least 3 times higher but were not tested for the lack of available power.

All our measurements were performed with photodiodes from one delivery. It shall be proved whether the behavior is similar for all traps with Hamamatsu S1337 1010.

---

[2] providing the change is lower than 10%/V, i.e. we are not in the very beginning of the curve in the fig.1

**References**

1. Fox N.P., *Metrologia*, 1991, **28**, 197-202.
2. Goebel R. and Stock M., *Metrologia*, 1998, **35**, 413-418.
3. Stock K.D., Morozova S., Liedquist L. and Hofer H., *Metrologia*, 1998, **35**, 451-454.
4. *Hamamatsu Photodiodes catalogue*, Hamamatsu Photonics K.K., 1998
5. Balling P., Fox N.P., Response Linearity Test of Biased 1337 Traps, In *Proc. CPEM 2000*, Sydney 2000, TUP6-2, 263-264